\documentclass{ws-ijmpcs}

\begin{document}

\markboth{C. Sigismondi}
{Solar Clementine gnomon's astrometric recognition}

%
\catchline{}{}{}{}{}
%
\title{THE ASTROMETRIC RECOGNITION OF THE SOLAR CLEMENTINE GNOMON (1702)}

\author{COSTANTINO SIGISMONDI}

\address{Sapienza University of Rome, Physics Dept., and Galileo Ferraris Institute\\
P.le Aldo Moro 5 Roma, 00185, Italy. e-mail: sigismondi@icra.it\\
University of Nice-Sophia Antipolis - Dept. Fizeau (France);\\
IRSOL, Istituto Ricerche Solari di Locarno (Switzerland)\\
}

\maketitle

\begin{history}
\received{06 Feb 2012}
\revised{Day Month Year}
\end{history}

\begin{abstract}

The Clementine gnomon has been built in 1702 to measure the Earth's obliquity variation. For this reason the pinhole was located in the walls of Diocletian's times (305 a. D.) in order to remain stable along the centuries, but its original form and position have been modified.
We used an astrometric method to recover the original position of the pinhole: reshaping the pinhole to a circle of 1.5 cm of diameter, the positions of the Northern and Southern limbs have been compared with the ephemerides. A sistematic shift of 4.5 mm Southward of the whole solar image shows that the original pinhole was 4.5 mm North of the actual position, as the images in the Bianchini's book (1703) suggest.
The oval shape of the actual pinhole is also wrong. Using a circle the larger solar spots are clearly visible. 
Some reference stars of the catalogue of Philippe de la Hire (1702), used originally for measuring the ecliptic latitude of the Sun, are written next to the meridian line, but after the last restauration (2000), four of them are wrongly located.   
Finally the deviation from the true North, of the meridian line's azimuth confirms the value recovered in 1750. This, with the local deviations of a true line, will remain as systematic error, like for all these historical instruments.
     
\end{abstract}

    
\ccode{PACS numbers: 01.65.+g; 96.60.-j; 96.60.Bn; 95.10.Jk}

\section{Introduction}    

All ancient meridian lines have been re-measured after some decades of duty, in order to verify their alignment and the position of the pinhole.
These instruments have been built to measure the variation of the obliquity along the centuries, and the need of a re-calibration was part of the observational duties.
The Cassini meridian line in San Petronio, Bologna, made in 1655 was revised in 1695 by the same astronomer Giandomenico Cassini.\cite{cassini}
Similarly Leonardo Ximenes in 1761 restaured the meridian line in Santa Maria del Fiore in Florence, made by Paolo del Pozzo Toscanelli in 1475.\cite{righini}
The great Clementine gnomon of Santa Maria degli Angeli in Rome, completed by Francesco Bianchini in 1702 was studied and remeasured by Anders Celsius in 1734 and Ruggero G. Boscovich in 1750.\cite{Heilbron}
They found the deviation of the azimuth from the true North, respectively of 2' (1734) and 4'30" (1750). 
Our measurements of 2006, used the Polaris' transits technique,\cite{monti} yielding $4'28".8\pm0.6"$, in agreement with the measurements made by Boscovich.\cite{sigi2010}

In the recognitions of Cassini and Ximenes the main issue was the movement of the pinhole with respect to the original position.

This was due to the fact that the pinhole in Bologna was on the roof, and in Florence was in the dome of the church: both positions were subjected to motions of the buildings due to thermal response, winds and settling of the walls.

For this reason Francesco Bianchini chosen the basilica of Santa Maria degli Angeli in Rome to build the meridian line upon the will of Pope Clement XI (1700-1721):
this church was built by Michelangelo in the original roman hall of the Diocletian baths, a 1500 years old structure, with no more settling ongoing.

The object of this paper is to show that the pinhole has been modified since 1700, and changed from its original form and position.
Therefore the measurements of the solar position and diameter at the meridian line are all affected by systematic errors, which can be easily corrected with a restauration of the original pinhole position and form.

\section{The Position of the Solar Center Recovered from Northern and Southern Limbs}

The pinhole, an horizontal hole which produces the solar image at local noon, is located in a thin metal frame that has been carved to obtain an elongated form, of $4\times 1.5$ cm with the major axis approximately in East-West direction.
For doing the measurements I have superimposed a circular mask of 1.5 cm of diameter in the center of the elongated pinhole.
In this way a sharper image can be observed on the floor of the church, and the solar spots NOAA 11401, 11402 and 11408 have been easily observed by several students and visitors (Fig. 1).

\begin{figure}
\centerline{\includegraphics[width=1\textwidth,clip=]{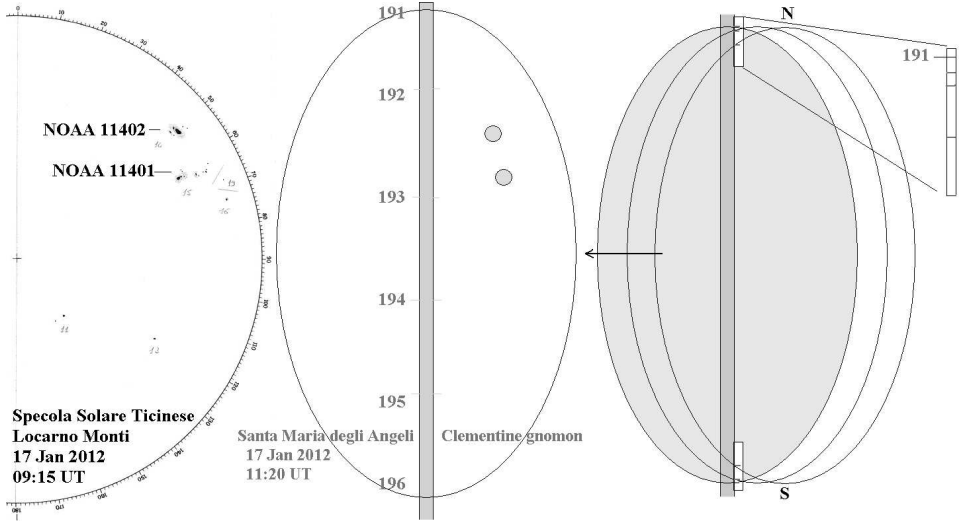}}
\caption{Sketch of the two solar spots NOAA 11401 and 11402 (umbra's diameter $\protect\emptyset=33$ arcsec) observed on the Clementine meridian line with the pinhole of $\protect\emptyset=1.5$ cm, at a focal distance F=44.3 m on 17 Jan 2012. The spots were perfectly visible on the marbles of the floor. The corresponding drawing of the Specola Solare Ticinese is reported on the left. 
The pinhole at that focal distance is 70 arcsec wide: the angular resolution. Dark umbras with $\protect\emptyset\ge 8$ arcsec like NOAA 11408 are still visible even if their contrast on the floor is $(8/70)^2$ times lower, because of the convolution with the pinhole spread function.\protect\cite{aip} Right: the progressive marks made on paper strips improve the measurement of North and South limbs' positions.}
\label{Fig. 1}
\end{figure}

Here the method and the data analysis to recover the solar center position during a meridian transit are presented.
Two strips of paper are prepared and posed near the meridian line, in correspondance with the positions of the Northern and the Southern limbs at the moment of the meridian transit. The strips are also referenced with respect to the original marks still existant on the lead meridian line.
For example the number 191 in Fig. 1 stands for $191=100\times tan(z_{\odot})$ and the corresponding zenithal angle is $z_{\odot}\sim 62.365^{\circ}$.  
At the same instants, slightly before and after the transit, the intersections between the solar ellipse on the floor and the paper strips are marked with a pencil, as in Fig. 1, right side. Southern and Northern limbs have to be marked in the same moment each time, therefore very rapidly or by two observers contemporarily, previously trained for this operation.
Southern and Northern marks taken at the same moments are simmetrically displaced with respect to the extreme positions attained by the solar limbs at the meridian transit. So the center of the Sun is recovered with the same algorithm either starting from the two extreme positions and from the other couples of intermediate positions.

\subsection{Differential Refraction}
The measurements for calibrating the position of the pinhole have been made also during the winter solstice of 2011.\cite{JOA}
The algorithm for taking into account of the phenomenon of atmospheric refraction which is slightly different from Southern and Northern limb is the following.
\begin{itemize}
\item{the Laplace formula $r=60"\times tan(z_{\odot})$ for the atmospheric refraction is enough precise for our purposes at height above the horizon $h_{\odot}\ge 24^\circ$.}
\item{the Northern and Southern limb $l_{N,\odot}, l_{S,\odot}$ positions without atmosphere are calculated starting from the center of the Sun  $z_{\odot}$ by adding and subtracting the solar radius  $r_{\odot}$.}
\item{the apparent positions of the limbs $l_{N,\odot,r}, l_{S,\odot,r}$ and of the center $z_{\odot,r}$ are calculated by including the atmospheric refraction with Laplace's formula.}
\item{the intermediate position between the two calculated (refracted) limbs is subtracted to the calculated central position
 $(l_{N,\odot,r} + l_{S,\odot,r})/2-z_{\odot,r} = \epsilon$.}
\item{$\epsilon$ is assumed constant for all simultaneous marks made with the pencil (they are within 4 arcmin from the limbs); 
$(l_{N,\odot,obs}+ l_{S,\odot,obs})/2 - \epsilon = z_{\odot,obs}$.}
\item{the various positions of $z_{\odot}$ recovered during the observational campaign are averaged to obtain the statistical dispersion.}
\item{the height of the pinhole is 20344 mm, measured in winter 2006 (2nd February) with modern theodolites, therefore the thermal expansion of the walls cannot be the responsible of the differences with respect to the ephemerides in the image's positions.}
\end{itemize}

\subsection{Results of the Displacement of the Pinhole}
The center of the present pinhole is $4.5\pm0.5$ mm South of the original position, as a result of the winter solstice campaign. 
At the winter solstice 0.5 mm on the meridian line correspond to 1 arcsec. This was the accuracy attained in 1700 with this instrument for measuring the center of the Sun.\cite{sigi2010} In a forthcoming restauration the pinhole has to be rebuilt 4.5 mm North of the present position. The original center of the pinhole corresponds to the metal window, as in the figure of Bianchini's book (1703).\cite{1703} The modification to the present position seems to be consequent with the definitive closing of the windows through which the Sun and the stars were observed contemporarily, in daytime. The modification of the pinhole shape to the present elongated form has been done before 1990, but not documented. This is completely wrong and has to be eliminated. The original shape was a circle, 1/1000 of the height of the pinhole, i. e. 2 cm. And only with this regular shape the solar spots are visible during at least 4 months around the winter solstice. To see the solar spots during summertime the pinhole has to be reduced to $\emptyset$= 7 mm.\cite{gnomone2009}

\begin{figure}
\centerline{\includegraphics[width=1\textwidth,clip=]{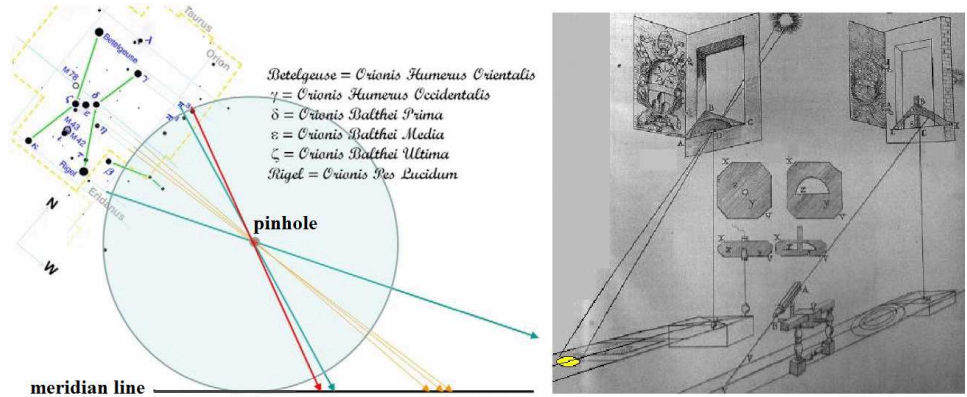}} 
\caption{The stars on the meridian line: scheme of the projection through the pinhole. Right: the window used to observe the Sun and the stars in daytime in 1700. The location of the solar pinhole was originally centered on the windows's border. Closing this window the pinhole has been moved externally (Southward).}
\label{Fig. 2}
\end{figure}

\section{Star's Names on the Meridian Line: Four Errors}
In order to remember the function of the meridian line to measure directly the ecliptic longitude $\lambda_{\odot}$ of the Sun as well as its declination $\delta_{\odot}$ on the meridian line there are the names of the reference stars with their ecliptic longitude. Their positions are from the Atlas of Philippe de la Hire (1701).\cite{hire} Presently, after the restauration of 2000 there are four errors,\cite{magda} three for changing sign to the declination read on the Atlas: Arietis lucida ($\alpha$ Ari, {\it Hamal}) that is represented at 201.7 instead of 36.1; Leonis Cauda ($\beta$ Leo, {\it Denebola}), which is at 125 instead of the correct 48.2, and Pegasi Os ($\epsilon$ Peg, {\it Enif}), which is at 119.5 instead of 65.8. Finally {\it Procyon}, Canis Minor ($\alpha$ CMi), has the ecliptic longitude of the following star in the Atlas: $110^o$ instead of the correct $104^o$ 1' 15".

\section{Conclusions}
The solar spots are clearly visible at this instrument. It is interesting to note that Paolo Toscanelli and his contemporaries did not report solar spots' observations because the meridian was made in 1475, during the Sp\"{o}rer minimum, lasted from 1460 to 1550.\cite{wei} Recently the transit of Venus was clearly observed with the Toscanelli gnomon in 2004. Therefore the missing solar spots at Toscanelli's gnomon in the first years of duty, confirms observationally the Sp\"{o}rer minimum 135 years before the introduction of the telescope.
When restaured, the circular shape of the pinhole in Santa Maria degli Angeli in Rome, will allow again the visibility of the largest solar spots.
The pinhole has to be replaced 4.5 mm Nortward, in the original position, and it will produce the images in the correct position with respect to the original marks, with an expected accuracy in the solar position determination of $\pm 1$ arcsec. The window used to observe the stars has to be restaured as well, otherwise the pinhole cannot be relocated in the correct position. Three stars {\it Hamal, Enif} and {\it Denebola} have been misplaced and they have to be moved in the Northern emisphere and the $\lambda$ of $\it Procyon$ has to be corrected.


\begin{thebibliography}{0}    

\bibitem{cassini} G. D. Cassini, {\it La Meridiana del tempio di S. Petronio}, {\bf },  Bologna (1695).

\bibitem{righini} F. Mazzucconi, P. Ranfagni, A. Righini, {\it Giornale di Astronomia}, {\bf 32}, 83 (2006).

\bibitem{Heilbron} J. L. Heilbron, {\it The Sun in the Church}, Harvard University Press (1999).

\bibitem{monti} C. Ferrari da Passano, C. Monti, L. Mussio, {\it La Meridiana Solare del Duomo di Milano, Verifica e Ripristino nell'anno 1976}, Milano (1977).

\bibitem{sigi2010} C. Sigismondi, {\it Astronomia UAI}, {\bf 3}, 56 (2011). {\rm http://arxiv.org/abs/1106.2498}

\bibitem{aip} C. Sigismondi, {\it American Journal of Physics}, {\bf 70}, 1157 (2002).  

\bibitem{JOA} C. Sigismondi,  {\it Journal of Occultation Astronomy}, in press (2012). {\rm http://arxiv.org/abs/1201.0510}

\bibitem{1703} F. Bianchini, {\it De Nummo et Gnomone Clementino}, {\bf },  Roma (1703).

\bibitem{gnomone2009} C. Sigismondi, {\it Lo Gnomone Clementino}, {\bf }, Roma (2009).

\bibitem{hire} Ph. de La Hire, {\it Tabulae Astronomicae Ludovici Magni}, {\bf } Paris (1702).

\bibitem{magda} M. Nastasi and M. Bedinsky, {\it Le Stelle e la Meridiana}, {\rm http://www.santamariadegliangeliroma.it/documenti/LESTELLEdellaMERIDIANA.pdf},  Roma (2007).

\bibitem{wei} W. Wei-Hock Soon, S. H. Yaskell, {\it The Maunder Minimum and the Variable Sun-Earth Connection}, World Scientific, Singapore (2003).

\end{thebibliography}
\end{document}